\documentstyle[twoside,fleqn,espcrc2,cite,epsfig]{article}

\newcommand{\bs}{\begin{slide}}
\newcommand{\es}{\end{slide}}
\newcommand{\be}{\begin{equation}}
\newcommand{\ee}{\end{equation}}
\newcommand{\bea}{\begin{eqnarray}}
\newcommand{\eea}{\end{eqnarray}}

\newcommand{\la}{\left\langle}
\newcommand{\ra}{\right\rangle}
\newcommand{\lc}{\left[}
\newcommand{\rc}{\right]}
\newcommand{\lp}{\left(}
\newcommand{\rp}{\right)}

\newcommand{\bc}{\begin{center}}
\newcommand{\ec}{\end{center}}

\newcommand{\red}{~}
\newcommand{\blue}{~}

\newcommand{\AmS}{{\protect\the\textfont2
  A\kern-.1667em\lower.5ex\hbox{M}\kern-.125emS}}
  \newcommand{\bi}{\begin{itemize}}
\newcommand{\ei}{\end{itemize}}
\def\epm#1#2{\hbox{${\lower1pt\hbox{$\scriptstyle +~#1$}}
\atop {\raise1pt\hbox{$\scriptstyle -~#2$}}$}}

\title{A probability measure in the space of spectral functions
and structure functions}

\author{Joan Rojo Chacon \address{Departament d'Estructura i 
Constituents de la Materia,
        Facultat de Fisica, \\ Universitat de Barcelona,
        Diagonal 647, 08028 Barcelona, Spain}}

\begin{document}

\begin{abstract}
We present a novel technique to parametrize experimental data,
based on the construction of a probability measure in the space of
functions, which retains the full experimental information on
errors and correlations. This measure is constructed in a two step
process: first, a Monte Carlo sample of replicas of the
experimental data is generated, and then an ensemble of neural
network is trained over them. This parametrization does not
introduce any bias due to the choice of a fixed functional form.
Two applications of this technique are presented. First a
probability measure in the space of the 
spectral function $\rho_{V-A}(s)$ is generated, which incorporates
theoretical constraints as chiral sum rules, and is used to
evaluate the vacuum condensates. Then we construct a probability
measure in the space of the proton structure function
$F_2^p(x,Q^2)$, which updates previous work, incorporating HERA data. 

\end{abstract}

\maketitle

\section{Introduction and motivation}
The general problem we are considering here\footnote{ Talk given
at the High-Energy Physics International Conference on Quantum
Chromodynamics at Montpellier (France), 5-9 July 2004. Based on the
work of Refs. \cite{Rojo:2004iq,f2neural2}.} is that of
the parametrization of experimental data. This is an ill-posed
problem, since it consists on obtaining continuous functions from
a finite set of measurements. Standard parameterizations, which
consist on choosing a functional form and fitting its parameters to the
data, have a series of shortcomings: first, the
choice of a functional form introduces an a priori bias, which
implies a theoretical uncertainty whose size is very difficult to
asses. Another important problem is how errors and correlations
are represented within this parametrization, and  how
uncertainties are propagated to other observables, since linear
error propagation is not trustable in general.

The motivation for this problem is the issue of Parton
Distributions Functions (PDFs) for the LHC \cite{qcdrev}.
Recently, a considerable amount of theoretical and experimental effort
has been invested in their accurate 
determination, and in particular their associate
errors, in view of the accurate computation of collider processes
and determination of QCD parameters. On the theory side, the PDFs
should be unbiased with respect to the choice of functional form, and
moreover
the full experimental information
should  be incorporated into the PDFs parametrization, including
systematic errors and correlations in a way that allows it to
propagate to observables (like cross sections) without introducing
an additional bias, linear approximations for instance. The technique
that we present here is specially devised to fulfill all these requirements.

\section{General strategy}
In this section we review the approach we take to this problem
\cite{Forte:2002fg}.
 The basic
idea is to
 construct a  probability measure in the space of functions, 
${\mathcal P}\lc f\rc$,
 from experimental information for $f$. From this
 probability measure one can compute any observable and the
 associated uncertainty and correlation,
 using weighted averages
\be
 \la {\mathcal  F}\lc f(\vec{x}) \rc\ra=\int {\mathcal D} f{\mathcal F}\lc
 f (\vec{x})\rc {\mathcal P} \lc  f \rc \ .
\ee
 The way  this idea is implemented in our formalism is using neural 
networks as universal unbiased
 interpolants, in a two step process. The first step is the  
Monte Carlo sampling, where we
generate a number of replicas of the artificial. 
Then we train an ensemble of neural networks
over these generated replicas. The whole procedure is then
validated through suitable statistical estimators.

The first step is the Monte Carlo sampling of experimental data,
which consists on the generation of $N_{rep}$ Monte Carlo sets of
'pseudo-data', replicas of the original $N_{dat}$ data points, $
f_i^{(art)(k)}$, $i=1,\ldots,N_{dat}$,$k=1,\ldots,N_{rep}$
 using equations
of the form \be  f_i^{(art)(k)} = \Delta_N\Big[ f_i^{(exp)}+r_i^s
\sigma^{stat}_i +\sum_{l=1}^{N_{sys}}r^{l,(k)}\sigma_i^{sys,l}
\Big] \ ,\ee where $\sigma_i^{stat}$ and $\sigma_i^{sys,l}$ are the
statistical and the different systematic
 errors of the point $i$ and 
$\Delta_N\equiv\lp 1+r_N^{(k)}\sigma_N\rp$ is the contribution from 
the normalization
 error. The $r_i$ are univariate gaussian random
 numbers. Correlated systematics share the same random
 numbers, and this takes into account the correlations in the
 generation of the replicas.

The second part of our technique consists on training one neural
network \cite{nn} on each Monte Carlo replica of the experimental
data\footnote{In Refs. \cite{Forte:2002fg ,Rojo:2004iq}
there is a detailed description of the neural networks and the
 learning algorithms used in this technique}. Neural networks are useful for
our purposes since they are the most unbiased prior, and they are
robust, unbiased universal approximants, so using them eliminates
the need of introducing a bias by choosing a functional form for
our parametrization.
 We use a combination of two different
techniques for training the networks: Backpropagation (BP)
learning and Genetic Algorithms (GA) learning. The second
technique, inspired in evolutionary models in in biology, has been
widely used in other branches of science. In this context training
means the minimization of an error function evaluated with the
covariance matrix for each replica \be
E^{(k)}=\sum_{i,j=1}^{N_{dat}}\Delta f_i^{(k)} {\mathrm cov}^{-1}_{ij}
\Delta f_j^{(k)} \ ,\ee where $\Delta f_i^{(k)}\equiv
f_i^{(art)(k)}-f_i^{(net)(k)} $, so that this error function
 measures  the goodness of the fit.  The set of trained nets is the
sought-for probability measure in the space of functions $f$, and
defines a parametrization of the experimental information for $f$.
Averages over this probability measure are performed using \be \la
{\mathcal F}\lc f_i\rc\ra=\frac{1}{N_{rep}}\sum_{k=1}^{N_{rep}}
{\mathcal F}\lc f^{(net)(k)}_i\rc
 \ .\ee
The final step consists on the validation of the fitting process using 
statistical estimators.

\section{Spectral functions}
Now we turn to consider the first of the two applications
\cite{Rojo:2004iq} of this technique. 
The vector and axial-vector spectral functions
$v_1(a),a_1(s)$ have been measured in hadronic tau decays at LEP
(ALEPH \cite{Barate:1998uf} and OPAL) up to $s=M_{\tau}^2$ with
large precision except near threshold. In particular we are
interested in parameterizing the vector-axial vector spectral
function $\blue \rho_{V-A}(s)\equiv v_1(s)-a_1(s)$. This spectral
function is interesting since it vanishes to all orders in
perturbation theory, and thus is specially suited to study
nonperturbative aspects of QCD. In particular this spectral
function is an order parameter of spontaneous chiral symmetry
breaking.

The combined use of the operator product
expansion and dispersion relations \cite{tau} allows an extraction
of the  QCD vacuum condensates in terms of convolutions of this
spectral function,
 \be \blue \la
{ \mathcal
O}_{2n+2}\ra=(-1)^n\int_0^{\infty}dss^n\frac{1}{2\pi^2}\rho_{V-A}(s) \ .
\ee
However, there are problems with experimental data to obtain a
clean extraction. First of all, we have information only up to the
tau mass kinematic threshold $s=M_{\tau}^2$, and second, there are
large errors and correlations near this threshold due to phase
space suppression factors. One possible solution consists on
constructing a probability measure in the space of spectral
functions $\rho_{V-A}(s)$ using the general technique introduced
above and then use the {\red chiral sum rules} to constrain the large
$s$ behavior within this probability measure.

Now we proceed as explained in Section 2. The only difference in
the GA training epoch consists on the minimization for each
replica of a modified error function,
 \be
 \label{mod}
 E^{(k)}+\sum_{l=1}^4g_l\lp \int_0^{\infty} ds
f_l(\rho_{V-A}^{(k)})-A_l\rp^2+E^{(k)}[s_0] \ee which takes into
account both the theoretical constraints from the chiral sum rules
(second term) and the asymptotic constraint
$\rho_{V-A}(s\to\infty)=0$ (third term). The chiral sum rules
\cite{sr} that are incorporated into the probability
measure are the Das-Mathur-Okubo
sum rule, the first and the second Weinberg sum rules (WSR), and
the electromagnetic mass splitting of the pion sum rule\footnote{It
 turns to be that the most relevant are the two
WSRs, $\int_0^{\infty}ds~\rho_{V-A}(s)=4\pi^2f_{\pi}^2 $ and
$\int_0^{\infty}ds~s\rho_{V-A}(s)=0 $ }. The use
of GA is crucial here since allows the learning of non-local error
functions like Eq. \ref{mod}.

So using this technique the probability measure for
$\rho_{V-A}(s)$ is constructed, and
from this measure we can evaluate any observable with the
corresponding uncertainty. The results for the lower dimensional
condensates are $
 \la {\mathcal O}_6 \ra=\lp -4.0 ~\pm
 2.0\rp~10^{-3}~\mathrm{GeV}^6$ and $
 \la {\mathcal O}_8 \ra=\lp -12 ~\epm{7}{11} \rp
10^{-3}~\mathrm{GeV}^8 $ (see Fig. 1). While the
result of $\la {\mathcal O}_6\ra$ is standard, the sign of $\la
{\mathcal O}_8 \ra$ oscillates in the 
literature\footnote{See Ref. \cite{Rojo:2004iq} for the details of 
this parametrization, a careful discussion on the sign of the dimension
8
condensate and a comparison with other determinations} \cite{condensates}.
We find
 that the uncertainties in the determination of the condensates  
are often underestimated and
dominated by  theoretical (model-dependent) uncertainties.

\begin{figure}[t]
\begin{center}
\epsfig{width=0.32\textwidth,figure=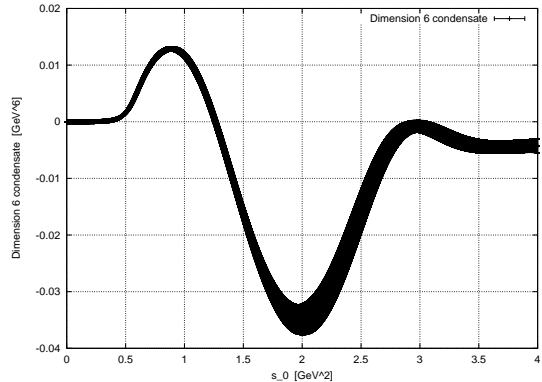,angle=-90}
\label{condensate2} \caption{The $\la {\mathcal O}_6\ra$ condensate
as a function of $s_0$. Error bands show propagation of
experimental uncertainties.}
\end{center}
\end{figure}

\section{Structure functions}
Now we present the second application \cite{f2neural2}, a
parametrization of the proton structure function $F_2^p(x,Q^2)$,
which is an update of Ref. \cite{Forte:2002fg}, with the following
novel features: incorporation of 11 more experiments (E665, H1 and
ZEUS, in addition to NMC and BCDMS),
 direct minimization of covariance matrix
error function and an improved analysis of experimental uncertainties (like a
dedicated treatment of asymmetric errors and uncorrelated
systematics). Note  that  a single fit  covers the whole kinematical range,
 $ 6~10^{-7}\le x \le 0.8$, $ 4.5~10^{-2}\le Q^2 \le
3~10^4 ~\mathrm{GeV}^2$, which consists on regions with very
different behaviour,  since we do not need to supply any
functional form.

\begin{figure}[t]
\begin{center}
\epsfig{width=0.32\textwidth,figure=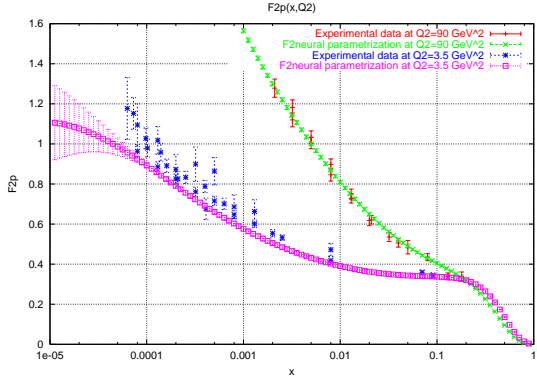,angle=-90}
\label{f2data2}\caption{Comparison of the F2neural fit with
experimental data. Only  diagonal
errors are shown, point-to-point correlations might be large.}
\end{center}
\end{figure}

\begin{figure}[t]
\begin{center}
\epsfig{width=0.32\textwidth,figure=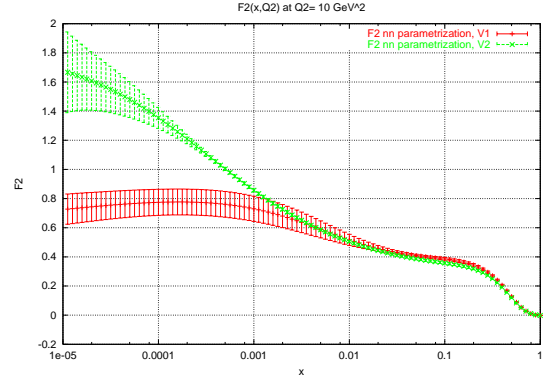,angle=-90}
\caption{Comparison of old and new versions of the F2neural fit. Note
  the characteristic increase of uncertainty in regions without
  experimental data}
\label{comparison}
\end{center}
\end{figure}

Following the steps of Section 2, we obtain a parametrization for
$F_2^p(x,Q^2)$. In Fig. 2 one
can observe our parametrization compared with experimental data.
Note that the uncertainties in $F_2^p(x,Q^2)$ are automatically
incorporated in the parametrization. Note also a relevant
feature: in regions without experimental data, the uncertainties
increases in a very characteristic way, so the region where the
parametrization ceases to be trustable is under control.

It is interesting to study the effect of the incorporation of
new experiments by comparing our parametrization with the old
version of F2neural \cite{Forte:2002fg}, where the only
experimental data was from the NMC and BCDMS experiments (see Fig.
\ref{comparison}). Note that the two fits are consistent in the
region with same experimental data (high $x$ region) but differ at
low $x$ where the new fit is  better, as was expected
since incorporates  the HERA data.

\section{Conclusions and future work}
We have presented a general technique to parametrize experimental
 data, applying it to two different problems.
 We have constructed a probability measure in the space of
spectral functions and structure functions showing that additional
theoretical constraints (sum rules, kinematical constraints) can
be  incorporated in the probability measure. These two 
applications increase the confidence on the validity of our approach.
 The next step is to construct a probability
measure in the space of PDFs, with full control over experimental
uncertainties, accurate error propagation and no bias due to the
choice of fixed functional forms.

\vspace*{0.5cm}

{\bf Acknowledgements}

We would like to thank the QCD04 Conference organizers. This work
has been supported by the projects MCYT FPA2001-3598,
GC2001SGR-00065 and by the Spanish
 grant AP2002-2415.

\end{document}